\documentclass[%
 reprint,
nofootinbib,
 amsmath,amssymb,
 aps,
]{revtex4-1}

\usepackage{graphicx}
\usepackage{dcolumn}
\usepackage{bm}
\usepackage[hidelinks]{hyperref}% add hypertext capabilities
\usepackage[mathlines]{lineno}% Enable numbering of text and display math
\usepackage[english]{babel}
\usepackage{amssymb}
\usepackage{amsmath}
\usepackage{latexsym}
\usepackage{cancel}
\usepackage{xcolor}
\usepackage{esint}

\usepackage[normalem]{ulem}
\def\lb{\label}
\def\be#1\ee{\begin{align}#1\end{align}}
\newcommand{\ppar}{{p_\parallel}}

\def\a{\alpha}
\def\b{\beta}

\def\lb{\label}

\newcommand{\appropto}{\mathrel{\vcenter{
  \offinterlineskip\halign{\hfil$##$\cr
    \propto\cr\noalign{\kern2pt}\sim\cr\noalign{\kern-2pt}}}}}

\begin{document}

\title{Unified description of galactic dynamics and the cosmological constant}

\author{Mariano Cadoni}
\email{mariano.cadoni@ca.infn.it}

\author{Andrea P. Sanna}
\email{asanna@dsf.unica.it}

\affiliation{Dipartimento di Fisica, Universit\`a di Cagliari, Cittadella Universitaria, 09042, Monserrato, Italy}
\affiliation{Istituto Nazionale di Fisica Nucleare (INFN), Sezione di Cagliari, Cittadella Universitaria, 09042 Monserrato, Italy}

\date{\today}

\begin{abstract}
We explore the phenomenology of a two-fluid cosmological model, where the field equations of general relativity (GR)  are sourced by baryonic and cold dark matter. We find that the model allows for a unified description of small and large scale, late-time cosmological dynamics. Specifically, in the static regime we recover the flattening of galactic rotation curves by requiring the matter density profile to scale as $1/r^2$. The same behavior describes matter inhomogeneities distribution at small cosmological scales.  This traces galactic dynamics back to structure formation. At large cosmological scales, we focus on back reaction effects of the spacetime geometry to the presence of matter inhomogeneities. We find that a cosmological constant with the observed order of magnitude, emerges by averaging the back reaction term on spatial scales of order $100 \ \text{Mpc}$ and it is related in a natural way to matter distribution. This provides a resolution to both the cosmological constant and the coincidence problems and shows the  existence of an intriguing link between the small and large scale  behavior in  cosmology. 
\end{abstract}

\maketitle

The $\Lambda$-Cold Dark Matter (CDM) model represents our current best understanding of the observed properties of the universe, by assuming that only $\sim 5 \%$ of its energy content is constituted by baryonic matter, while the remaining $\sim 95 \%$ is exotic. Specifically, $\sim 30 \%$ is associated to non-baryonic CDM, $\sim 65 \%$ to dark energy in the form of a cosmological constant (CC) $\Lambda$ \cite{Aghanim:2018eyx}. The former allows for a simple explanation of a wide variety of observations, ranging from the flattening of rotation curves in disk galaxies and the internal dynamics of galaxy clusters to cosmological structure formation and evolution, the abundances of light elements and the power spectrum of the cosmic microwave background radiation \cite{Aghanim:2018eyx, Bertone:2004pz, Cuoco:2003cu}. The CC accounts 
instead for the observed accelerated expansion of the universe \cite{Riess:1998cb, Perlmutter:1998np}.\\
Despite these successes, several questions remain open. The most pressing ones are perhaps the understanding of the nature of the dark components, the explanation of the origin of galactic and cluster dynamics in terms of their formation and  explaining why the CC has the observed value and why its energy density is so closed to that of matter in the present epoch   (the cosmological constant  and the  coincidence problems)\cite{Peebles:2002gy}. \\
In this letter, we build on the results of Ref. \cite{Cadoni:2020jxe}, which are sufficiently general to be applied to various scenarios where cosmology is sourced by a two-fluid system. Working in the standard $\Lambda$CDM  framework and using baryonic matter and CDM (whose existence is here assumed) as sources of the gravitational field, we tackle some of the aforementioned problems of  late-time cosmology. Specifically, our model reproduces, in the static regime, the flattening of galactic rotation curves, whereas at  small cosmological scales explains local inhomogeneities and structure formation. This last result, in particular, traces the origin of galactic dynamics back to structure formation. At large cosmological scales, back reaction effects of the geometry to the presence of matter inhomogeneities are investigated. It is shown that, when averaged on spatial scales of order $100 \ \text{Mpc}$, they reproduce an effective cosmological constant, whose order of magnitude agrees with observations (this solves the CC problem). The origin of $\Lambda$ is thus linked to matter distribution, which solves the coincidence problem.

\textit{The model}.---Our model of late-time cosmology is GR sourced by a two-fluid system, consisting of baryonic and cold dark matter. We adopt the standard description and we model them as two pressureless perfect fluids, interacting only gravitationally one with each other, with densities $\rho_B$ and $\rho_{DM}$ and 4-velocities $U_{\mu}$ and $W_{\mu}$ respectively. The stress-energy tensor is then :
\begin{equation}
T_{\mu\nu} = \rho_B U_{\mu}U_{\nu}+ \rho_{DM} W_{\mu}W_{\nu}.
\label{EItwofluids}
\end{equation}
It is known that Eq.~(\ref{EItwofluids}) can be recast as the stress-energy tensor of an anisotropic fluid by an appropriate rotation of $U_{\mu}$ and $W_{\mu}$ \cite{Bayin:1985cd}. In the current  case, this transformation yields
\begin{equation}
T_{\mu\nu} = \rho u_{\mu}u_{\nu} + p_{\parallel}w_{\mu}w_{\nu},
\label{SETDMBM}
\end{equation}
describing an anisotropic fluid with zero tangential pressure $p_{\perp}$. $\rho\simeq \rho_B+\rho_{DM}$  (for $W^{\mu}U_{\mu}\simeq 1$), $u_{\mu}u^{\mu}=-w_{\nu}w^{\nu}=-1$ and $\ppar$ can be interpreted as an effective radial pressure stabilizing the  dark matter  halo. This is conceptually equivalent to the description of a fluid of collisionless particles, where an effective pressure term can be associated to the stress tensor modeling the anisotropy of the velocity distributions \cite{Jeans22, Binney:1982jf, Herrera:1997plx}. This is particularly suited for dark matter, which is believed to be made of collisionless particles. The two-fluid approach has the advantage to provide such a description in a natural and  straightforward way. 

To describe both the galactic and  cosmological regime, we use the following general, spherically symmetric, spacetime metric (we use units with the speed of light $c=1$)
\begin{equation}
ds^2 = a^2(t)\left[-e^{\alpha(t, r)} dt^2 + e^{\beta(t, r)} dr^2 + r^2 d\Omega^2\right],
\label{metrica}
\end{equation}
where  $\alpha(t, r)$ and $\beta(t, r)$ are metric functions, $a$ is the cosmological scale factor and $d\Omega^2 = d\theta^2 + \sin^2 \theta d\phi^2$. \\
The resulting independent Einstein field and conservation equations are (we use  $'=\partial_r,\, \dot\,  = \partial_t$):
\begin{equation}
3\frac{\dot{a}^2}{a^2}e^{-\a}+\frac{e^{-\beta}}{r^2}\left(-1+e^{\beta}+r\beta' \right)+\frac{\dot{a}}{a}\dot{\beta} e^{-\a}= 8\pi G a^2 \rho;
\label{E00}
\end{equation}

\begin{equation}
\frac{\dot{a}}{a}\alpha'+\frac{\dot{\beta}}{r}=0;
\label{E0r}
\end{equation}

\begin{equation}
\frac{\left(1+r\a'\right)-e^{\b}}{e^{\b}a^2r^2}-e^{-\alpha} \left(2\frac{\ddot{a}}{a^3}-\frac{\dot{a}}{a^3}\dot{\alpha} -\frac{\dot{a}^2}{a^4}\right)= 8\pi G \ppar
\label{Err}
\end{equation}

\begin{equation}
\dot{\rho}+\frac{\dot{a}}{a}\left(3\rho+\ppar\right)+\frac{\dot{\beta}}{2}\left(\rho+\ppar\right)=0;
\label{T0}
\end{equation}

\begin{equation}
\ppar'+\frac{\alpha'}{2}\left(\rho+\ppar\right)+\frac{2}{r}\ppar=0.
\label{Tr}
\end{equation}
In particular, in the static case, Eq.~(\ref{Tr}) is the generalization of the Newtonian hydrostatic equilibrium equation
\begin{equation}
p'  = -\frac{\partial_r  e^{\a}}{2}  \rho=-  \Phi' \rho= - \frac{G m(r)\rho}{r^2},
\label{HydroeqNew}
\end{equation}
where $\Phi(r)$ is the gravitational potential. In fact, using the weak field result  $e^\alpha\simeq 1+2\Phi(r)$,  Eq.~(\ref{Tr}) gives the Tolmann-Oppenheimer-Volkoff-like equation:
\begin{equation}
\partial_r \left(r^2  \ppar \right) = -r^2\partial_r \Phi \left(  \rho+ \ppar \right). 
\label{Hydroeq1}
\end{equation}

\textit{Galactic regime}.--- The galactic regime is obtained  by taking the static limit of Eqs.~(\ref{E00})-(\ref{Tr}), which implies the scale factor being constant (we set $a=1$), and the metric functions depending  on $r$ only. Integration of Eq.~(\ref{E00}) in this case gives %
\begin{equation}\lb{MisShmassgalactic}
e^{-\beta} = 1-\frac{2GM(r)}{r},
\end{equation}
where $M(r) \equiv 4\pi \int \rho(r) r^2  dr$ is the Misner-Sharp (MS) mass of the system. The remaining equations are solved together with a given profile for the matter density  $\rho =\rho(r)$.
We  are looking for solutions  reproducing the flattening of rotation curves at galactic scales. This can be achieved by choosing the following 
matter density profile% 
\begin{equation}
\rho= \frac{\sigma}{r^2},
\label{staticrho}
\end{equation}
with $\sigma$ a constant. This gives the MS mass $M(r) = 4\pi\sigma r $. In fact, virializing the galactic motion, we get the velocities $v^2=8\pi G\sigma$. We skip a  constant term in $M(r)$, which represents  the contribution of  the mass contained in the central regions of the galaxy. Here, we are considering only galactic scales $\gg \text{kpc}$, where, according to observations (see e.g. Refs. \cite{Rubin:1980zd, Bosma:1981zz}), rotation curves starts flattening. \\
With the density profile~(\ref{staticrho}), the field equations can be integrated to give
\begin{subequations}
\begin{align}
&\a = 2 \ln \left[\mathcal{C} \ln \left(\frac{r}{L} \right) \right]; \lb{alphastaticsol}\\
&\ppar =  -\frac{\sigma}{r^2}+\frac{1-8\pi G \sigma}{4\pi G}\frac{1}{r^2 \ln \left(r/L\right)}, \label{pstaticsol}
\end{align}
\end{subequations}
with $\mathcal{C}$ and $L$ integration constants. These are galactic parameters and could be determined, together with $\sigma$, by combining rotation curves and gravitational lensing (see, e.g. Ref. \cite{Faber:2005xc}).  \\
We note that
\begin{equation}
\ppar=-\rho=-\frac{\sigma}{r^2},\quad \a=0  
\label{sol2}
\end{equation}
also solves the field equations. This solution, giving an equation of state (EOS) $\ppar=-\rho$ and a \textit{negative} pressure, dominates for $r\to \infty$, i.e in the transition  to the cosmological regime.
Conversely, the second, \textit{positive}, term in Eq. (\ref{pstaticsol}) dominates at smaller (galactic) scales. Physically, this means that, in this regime, the hydrostatic equilibrium of dark matter halos is obtained  by contrasting the gravitational pull with a positive radial pressure. On the other hand, at large distances, in the transition to the cosmological  regime, the hydrostatic  equilibrium is  not reached in the usual intuitive way.  It is a local equilibrium  in which  both sides of Eq. (\ref{Hydroeq1}) separately vanish. This is  possible only if the EOS is $\ppar=-\rho$ and the pressure is negative.  As it is already evident from the form of the EOS, this is strongly related to  the generation of the cosmological constant in the cosmological regime (see below).\\
The existence of solution  (\ref{sol2})  is a peculiar feature of fluids with anisotropies.  Static isotropic fluids  do  not allow for solutions with $\a'=0$ and  $p,\,\rho$ satisfying $p=-\rho$.

Our static solution, describing hydrostatic equilibrium of the DM medium,  models a spacetime with a conical singularity (the relevance of this kind of solution for DM has been already noted in  \cite{Muckprivate}). In the simplest case, given by Eq.~(\ref{sol2}), the metric is:
\begin{equation}
ds=-dt^2+ \frac{dr^2}{(1-8\pi G \sigma)}+r^2d\Omega^2.  
\label{metric}
\end{equation}
The solution is physically  acceptable in the weak field limit when the the deficit angle is very small:
  \begin{equation}
 \sigma\ll \frac{1}{8\pi G}. 
\label{wf}
\end{equation} 
 In this limit, the spacetime can be well approximated by flat Minkowski space. The flattening of galactic  rotation curves is observed  when the acceleration drops below $a_0\sim \ell^{-1}$ (with $\ell$ the size of the cosmological horizon).
This implies that  condition (\ref{wf}) is satisfied for $r\ll l$,  which covers not only galactic, but also larger scales where our universe appears inhomogeneous, with the density of inhomogeneities scaling as $1/r^2$ \cite{Cadoni:2020jxe}.
The behavior  $\rho \sim 1/r^2$ is thus responsible not only for the flattening of galactic rotation curves, but also well describes  structure distribution at small cosmological scales \cite{Cadoni:2020izk}. This  means, physically, that the dynamical properties of galaxies are inherited from structure formation.\\
 The condition $r\ll l$ holds true also at scales $\mathcal{R}\sim 100 \ \text{Mpc}$, where our universe begins to appear homogeneous and isotropic. It breaks down for $r\sim \ell$, i.e.  in the cosmological regime where the static approximation is no longer valid and we have to consider the full form of the metric (\ref{metrica}).

 \textit{Cosmological regime}.---When $a \neq 1$ and the assumption of staticity of the metric functions is dropped, the system of equations~(\ref{E00})-(\ref{Tr}) describes the cosmological regime of our model. The matter density  determines  the  metric function  $\beta(t,r)$, whereas the back reaction of the metric  to the presence of matter is codified in the function $\a(t,r)\neq \text{constant}$ \cite{Cadoni:2020jxe}. In the decoupling limit, when the back reaction   can be neglected,  the cosmological dynamics is described by the standard Friedmann-Lemaitre-Robertson-Walker (FLRW) cosmology and  decouples completely from inhomogeneities  \cite{Cadoni:2020izk}. 
 
In the cosmological regime, exact solutions of the field equations (\ref{E00})-(\ref{Tr}) can be found using a method similar to that used in Ref. \cite{Cadoni:2020jxe}. One first  integrates  Eq.~(\ref{E0r}), defines the rescaled quantities 
$\hat{\rho} \equiv 3e^{\a}(3-r\a')^{-1} \rho,\, \hat{p}_{\parallel} \equiv e^{\a}\ppar$ and  then  uses an ansatz to separate the standard FLRW dynamics from that of inhomogenities and the back reaction:
\begin{align}
&a^2 \hat{\rho}(t,r) \equiv  a^2{\rho}^{(1)}(t)+\frac{3 e^{\alpha}}{3-r\alpha'}\left(\rho^{(2)}(r)+ \rho^{(3)}(t, r)\right), \nonumber\\
&a^2 \hat{p}_{\parallel}(t,r) \equiv  a^2{p}_{\parallel}^{(1)}(t)+ e^{\alpha}\left({p}_{\parallel}^{(2)}(r)+ {p}_{\parallel}^{(3)}(t, r)\right)\label{hat1}. 
\end{align}
The system (\ref{E00})-(\ref{Tr}) splits in the  usual FLRW equations  for $a$, sourced by ${\rho}^{(1)}$ and  ${p}_{\parallel}^{(1)}$, together with  the solutions for $\beta$ and $\a$ (see Ref. \cite{Cadoni:2020jxe} for the calculation details) .
The solutions  for  $\beta$ is also here given by  Eq.~(\ref{MisShmassgalactic}),  with the MS mass  being $M(t, r) \equiv M^{(1)}(t)+M^{(2)}(r) + M^{(3)}(t, r)$, where $M^{(1)}$ is an integration function, while $M^{(2,3)}$ are the MS masses associated to $\rho^{(2,3)}$ respectively.
Finally, the   solution  for $\a$ turns out to be 
\be\label{a3p} 
\a(t,r) = \mathcal{A}(t) + 2G \ \frac{a}{\dot{a}}\int \frac{\dot{M}}{r^2}\left(1-\frac{2GM}{r} \right)dr,
\ee
with $\mathcal{A}(t)$ integration function. We note that this solution, and in particular the  $(r,t)$-dependent terms, in general,  removes the conical singularity of the static solution.

The next step is to describe cosmology near the transition scale $\mathcal{R}$ to homogeneity and isotropy. In this situation, we cannot simply use the general solution written above, since it describes also inhomogeneities and their interaction with the cosmological dynamics encoded in the scale factor $a$.  Also the decoupling limit  does not seem appropriate because it completely neglects the back reaction. The simplest way to circumvent this  problem is  first to split $\a$ into functions depending only on $r$ and $t$, i.e. $\a \equiv \a_r(r) + \a_t(t)$. Then, we expand the solutions near the decoupling limit, i.e. $r\a_r' =0$, and near the present epoch of our universe, i.e. $a^2 =1$. Finally, we perform the spatial average of the resulting $r$-dependent quantities. Keeping only the leading  terms in the expansions, Eqs.~(\ref{hat1}) give  ($\a_t$ can be absorbed by a rescaling of $t$)
\begin{align}
&a^2 \rho(t) =  \frac{3}{8\pi G} \left(\frac{\dot{a}}{a} \right)^2+ a^2 \langle\rho^{(3)}\rangle_r\label{alpha006};\\
&a^2 p(t) =\frac{1}{8\pi G}\left[\left(\frac{\dot{a}}{a} \right)^2-2\frac{\ddot{a}}{a}\right]
 - a^2 \Bigg\langle\frac{M^{(3)}}{4\pi r^3}\Bigg\rangle_r\label{alpharr6},
\end{align}
where $\rho^{(3)}$, for consistency reasons, is function of $r$ only and the spatial averaging is performed  on spatial scales of order $100 \ \text{Mpc}$   (see \cite{Cadoni:2020jxe} for further details). 
When  $\rho^{(3)}\sim 1/r^2$, these equations  describe  standard FLRW cosmology with the averaged back reaction term $\langle\rho^{(3)}\rangle_r$ playing the  role of a cosmological constant, 
\be \lb{Lambdaaverage}
\Lambda \sim -8\pi G \langle \rho^{(3)}\rangle_r, 
\ee
corresponding to a perfect fluid  with equation of state $p=-\rho$.  It is important to stress that spatial averaging at length scales of order $100 \ \text{Mpc}$ solves also the conical singularity issue. In fact,  the weak field condition  (\ref{wf}) is satisfied and the $t=\text{constant}$ sections  of our metric are regular,  flat Minkowski spacetime. Notice  that  the emergence of a cosmological constant at large scales could  have been guessed   directly from  the existence  of  the  static solution  (\ref{sol2}), which dominates at large $r$.  

Let us now evaluate the order of magnitude of the cosmological constant.
In  Ref. \cite{Cadoni:2020jxe} it is argued that $\rho^{(3)} \sim \rho^{(2)}$, since $\rho^{(3)}$ is the energy density of the back reaction of the geometry to the presence of matter inhomogeneities, described by $\rho^{(2)}$. As we have previously seen, modelling  both dark matter at galactic scales  and the structure formation at small cosmological scales   \cite{Cadoni:2020izk,Cadoni:2020jxe} is consistent with  $\rho^{(2)}$ scaling as $1/r^2$. We have therefore $\rho^{(3)} \appropto -\frac{1}{r^2}$,  
where the minus sign is due to the fact that $\rho^{(3)}$ should give rise to an attractive force and must behave as an inverse power of $r$, hence $\Lambda \sim 8\pi G \langle \rho^{(2)}\rangle_r$.
The spatial average  can be easily computed in terms of the total mass $M$ of dark matter inside a sphere of radius $R$  of order  $100 \ \text{Mpc}$:  $\langle \rho^{(2)}\rangle_r=3M/4\pi R$. We have   $M \sim 10^{18} \ M_{\odot}$  and    $\Omega_{\Lambda0}\sim 1$, giving  the correct order of magnitude of the observed cosmological constant \cite{Aghanim:2018eyx}.  The fact that the energy densities associated to matter and $\Lambda$ are of the same order of magnitude at the present epoch does not appear here as a coincidence. This solves the coincidence problem of  the standard $\Lambda$CDM cosmological scenario.
\\

\textit{Conclusions}.---In the present work, we have presented and explored the phenomenology of a cosmological model sourced by baryonic and cold dark matter in late-time cosmology. Our model allows for a unified description of  galactic dynamics and cosmological  dynamics as well as  structure formation. 
The flattening of the rotation curves of galaxies,  the formation and distribution of structures at small cosmological scales and the cosmological constant all  have the same origin in the distribution of matter inhomogeneities and in back reaction of  the geometry to the presence of the latter.  The observed galactic dynamics  and structure formation and distribution are correctly reproduced by assuming the presence of an anisotropic component of the pressure and  of an $1/r^2$ scaling of matter density.  The observed  order of magnitude of the cosmological constant is explained as the average of the back reaction of the geometry at  scales where our universe starts appearing homogeneous and isotropic.    
The results of our  work  show the  existence of an intriguing link between the small and large scale  behavior in cosmology. Hints on this direction come also from the  presence of the fundamental acceleration scale $a_0$ in the  baryonic Tully-Fisher relation, which is of the order of magnitude of the present Hubble acceleration \cite{McGaugh:2004aw}.


\begin{thebibliography}{100}
\bibitem{Aghanim:2018eyx}
N.~Aghanim \textit{et al.} [Planck],
``Planck 2018 results. VI. Cosmological parameters,''
\href{http://dx.doi.org/10.1051/0004-6361/201833910}
{Astron. Astrophys. \textbf{641} (2020), A6},
\href{https://arxiv.org/abs/1807.06209}{{\ttfamily arXiv:1807.06209 [astro-ph.CO]}}.

\bibitem{Bertone:2004pz}
G.~Bertone, D.~Hooper and J.~Silk,
``Particle dark matter: Evidence, candidates and constraints,''
\href{http://dx.doi.org/10.1016/j.physrep.2004.08.031}
{Phys. Rept. \textbf{405} (2005), 279-390},
\href{https://arxiv.org/abs/hep-ph/0404175}{{\ttfamily arXiv:hep-ph/0404175 [hep-ph]}}.

\bibitem{Cuoco:2003cu}
A.~Cuoco, F.~Iocco, G.~Mangano, G.~Miele, O.~Pisanti and P.~D.~Serpico,
``Present status of primordial nucleosynthesis after WMAP: results from a new BBN code,''
\href{http://dx.doi.org/10.1142/S0217751X04019548}
{Int. J. Mod. Phys. A \textbf{19} (2004), 4431-4454},
\href{https://arxiv.org/abs/astro-ph/0307213}{{\ttfamily arXiv:astro-ph/0307213 [astro-ph]}}.


\bibitem{Riess:1998cb}
 A.~G.~Riess {\it et al.} [Supernova Search Team],
  ``{Observational evidence from supernovae for an accelerating universe and a
  cosmological constant},'' \href{http://dx.doi.org/10.1086/300499}{{\em
  Astron. J.} {\bfseries 116} (1998) 1009--1038},
\href{http://arxiv.org/abs/astro-ph/9805201}{{\ttfamily arXiv:astro-ph/9805201
  [astro-ph]}}.
  
\bibitem{Perlmutter:1998np}
  S.~Perlmutter {\it et al.} [Supernova Cosmology Project Collaboration],
  ``Measurements of $\Omega$ and $\Lambda$ from 42 high redshift supernovae,''
  Astrophys.\ J.\  {\bf 517} (1999) 565,
  \href{https://arxiv.org/abs/astro-ph/9812133}{{\ttfamily arXiv:astro-ph/9812133v1}}.

\bibitem{Peebles:2002gy}
P.~J.~E.~Peebles and B.~Ratra,
``The Cosmological Constant and Dark Energy,''
\href{http://dx.doi.org/10.1103/RevModPhys.75.559}
{Rev. Mod. Phys. \textbf{75} (2003), 559-606},
\href{https://arxiv.org/abs/astro-ph/0207347}{{\ttfamily arXiv:astro-ph/0207347 [astro-ph]}}.

\bibitem{Cadoni:2020jxe}
M.~Cadoni and A.~P.~Sanna,
``Emergence of a Cosmological Constant in Anisotropic Fluid Cosmology,''
\href{https://arxiv.org/abs/2012.08335}{{\ttfamily arXiv:2012.08335 [gr-qc]}}.

  \bibitem{Bayin:1985cd}
S.~S.~Bayin,
``Anisotropic fluids and cosmology,''
 \href{http://dx.doi.org/10.1086/164056}
{Astrophys. J. \textbf{303} (1986), 101-110}.


 \bibitem{Jeans22}
J.~H.~Jeans,
``The Motions of Stars in a Kapteyn Universe,''
\href{http://dx.doi.org/10.1093/mnras/82.3.122}
{Mon. Not. R. Astron. Soc. \textbf{82} (1922), 122-132}.

\bibitem{Binney:1982jf}
J.~Binney,
``Dynamics of elliptical galaxies and other spheroidal components,''
\href{http://dx.doi.org/10.1146/annurev.aa.20.090182.002151}
{Ann. Rev. Astron. Astrophys. \textbf{20} (1982), 399-429}.

\bibitem{Herrera:1997plx}
L.~Herrera and N.~O.~Santos,
``Local anisotropy in self-gravitating systems,''
\href{http://dx.doi.org/10.1016/S0370-1573(96)00042-7}
{Phys. Rept. \textbf{286} (1997), 53-130}.

\bibitem{Rubin:1980zd}
V.~C.~Rubin, N.~Thonnard and W.~K.~Ford, Jr.,
``Rotational properties of 21 SC galaxies with a large range of luminosities and radii, from NGC 4605 (R = 4kpc) to UGC 2885 (R = 122 kpc),''
\href{http://dx.doi.org/10.1086/158003}
{Astrophys. J. \textbf{238} (1980), 471}.


\bibitem{Bosma:1981zz}
A.~Bosma,
``21-cm line studies of spiral galaxies. 2. The distribution and kinematics of neutral hydrogen in spiral galaxies of various morphological types,''
\href{http://dx.doi.org/10.1086/113063}
{Astron. J. \textbf{86} (1981), 1825}.


\bibitem{Faber:2005xc}
T.~Faber and M.~Visser,
``Combining rotation curves and gravitational lensing: How to measure the equation of state of dark matter in the galactic halo,''
\href{http://dx.doi.org/10.1111/j.1365-2966.2006.10845.x}
{Mon. Not. Roy. Astron. Soc. \textbf{372} (2006), 136-142},
\href{https://arxiv.org/abs/astro-ph/0512213}{{\ttfamily arXiv:astro-ph/0512213 [astro-ph]}}.

\bibitem{Muckprivate}
W.~M\"uck, 
Private communication.


\bibitem{Cadoni:2020izk}
M.~Cadoni, A.~P.~Sanna and M.~Tuveri,
``Anisotropic fluid cosmology: An alternative to dark matter?,''
\href{http://dx.doi.org/10.1103/PhysRevD.102.023514}
{Phys. Rev. D \textbf{102} (2020) no.2, 023514},
\href{https://arxiv.org/abs/2002.06988}{{\ttfamily arXiv:2002.06988 [gr-qc]}}.

\bibitem{McGaugh:2004aw}
S.~S.~McGaugh,
``The Mass discrepancy - acceleration relation: Disk mass and the dark matter distribution,''
\href{http://dx.doi.org/10.1086/421338}
{Astrophys. J. \textbf{609} (2004), 652-666},
\href{https://arxiv.org/abs/astro-ph/0403610}{{\ttfamily arXiv:astro-ph/0403610 [astro-ph]}}.


\end{thebibliography}
\end{document}